\documentclass[aps,prb,floats,floatfix,twocolumn,longbibliography,superscriptaddress]{revtex4-2}

\usepackage{newtxtext}
\usepackage{newtxmath}
\usepackage{amsmath}
\usepackage{color}
\usepackage{graphicx}
\usepackage{epsfig}
\usepackage{amsfonts}
\usepackage{mathrsfs}
\usepackage{mathtools}
\usepackage{amsmath}
\usepackage{float}
\usepackage{physics}
\usepackage{hyperref}
\hypersetup{
	colorlinks=true,
	citecolor=blue,
	linkcolor=blue,
	urlcolor=blue}

\DeclareSymbolFont{myletters}{OML}{ztmcm}{m}{it}
\DeclareMathSymbol{\uplambda}{\mathord}{myletters}{"15}

\DeclareSymbolFont{cmlargesymbols}{OMX}{cmex}{m}{n}
\let\sum\relax
\DeclareMathSymbol{\sum}{\mathop}{cmlargesymbols}{"50}
\DeclareSymbolFont{cmletters}{OML}{cmm}{m}{it}
\SetSymbolFont{cmletters}{bold}{OML}{cmm}{b}{it}
\DeclareSymbolFontAlphabet{\mathnormal}{cmletters}

\begin{document}
\title{Spin-orbit coupled spin-boson model : A variational analysis}

\author{Sudip Sinha}
\affiliation{School of Physics, Korea Institute for Advanced Study, Seoul 02455, Korea}
\author{Subhasis Sinha}
\affiliation{Indian Institute of Science Education and Research-Kolkata, Mohanpur, Nadia-741246, India}
\author{Sushanta Dattagupta}
\affiliation{Sister Nivedita University, New Town, Kolkata 700156, India}

\begin{abstract}
The spin-boson (SB) model is a standard prototype for quantum dissipation, which we generalize in this work, to explore the dissipative effects on a one-dimensional spin-orbit (SO) coupled particle in the presence of a sub-ohmic bath. We analyze this model by extending the well-known variational polaron approach, revealing a localization transition accompanied by an intriguing change in the spectrum, for which the doubly degenerate minima evolves to a single minimum at zero momentum as the system-bath coupling increases. For translational invariant system with conserved momentum, a continuous magnetization transition occurs, whereas the ground state changes discontinuously. 
We further investigate the transition of the ground state in the presence of harmonic confinement, which effectively models a quantum dot–like nanostructure under the influence of the environment.
In both the scenarios, the entanglement entropy of the spin-sector can serve as a marker for these transitions. Interestingly, for the trapped system, a cat-like superposition state corresponds to maximum entanglement entropy below the transition, highlighting the relevance of the present model for studying the effect of decoherence on intra-particle entanglement in the context of quantum information processing.
\end{abstract}

\date{\today}
\maketitle
\section{Introduction}\label{1}
The paradigmatic spin-boson (SB) model provides the simplest framework to understand the rich variety of phenomena arising from quantum dissipation, with applications ranging from physics to biological systems \cite{Petruccione_book, Weiss_book, Leggett_review}. This model describes the non-equilibrium dynamics of a two level system or equivalently a spin-$1/2$ particle coupled to a bosonic bath. Various dynamical regimes can be identified by varying the coupling to the bath \cite{Leggett_review}. Importantly, if the two spin states correspond to the two minima of the symmetric double well, the SB model exhibits a localization transition for sufficiently strong system-bath coupling strength \cite{Weiss_book, Leggett_review, Leggett_Chakravarty}, representing the suppression of tunneling due to the influence of the environment \cite{Leggett_Chakravarty, Caldeira_Leggett_1981, Bray_Moore_1982}. This localization transition has been extensively studied using various methods \cite{Silbey_Harris, Vojta2003, Bulla2005, Bulla2008, Anders2007, Vojta2005, Vojta2009, Cohen2025, Fehske2009, Zhang2010, Hou2010, Chin2006, plenio_prl, Zheng2011, Tong2011,  Chin2014_environmental_entanglement, Chen2018, Naushad2024, NaushadPRX}.
Moreover, the critical behavior of this transition crucially depends on the spectral properties of the bath \cite{Leggett_review, Bulla2005, Bulla2008, Vojta2005, Hou2010}. Interestingly, a localization transition induces an effective asymmetry in an otherwise symmetric double well, leading to a continuous magnetization transition in the SB model with a sub-ohmic bath, which can be described by a Landau-Ginzburg theory \cite{plenio_prl}. 
In recent years, the SB model has regained significant interest due to its crucial role in quantum information theory. It effectively captures the dissipative dynamics of a qubit interacting with its environment, making it a valuable tool for studying decoherence and quantum measurement processes \cite{Schlosshauer2019, Unruh1995, Shnirman2002, Hanggi2001, Wilhelm2003, Khveshchenko2004, Hur2008, Kamil2012, Fischer2014}. Furthermore, system-bath entanglement is a critical quantity that has been employed to investigate the localization transition within the SB model \cite{plenio_prl, Zheng2011, Hur2008, McKenzie2003, Hur2007, Zhou2022}.

In many physical systems, the dynamics of a (quasi)particle are often linked to its spin degrees of freedom, especially in the context of spin-orbit (SO) coupling \cite{SOC_Dresselhaus, SOC_Rashba, SOC_Rashba_review}, and in materials governed by effective Dirac Hamiltonians, such as Graphene and topological insulators \cite{Graphene_review,Graphene_review2,Topological_insulators_colloquim}.
Furthermore, recent experimental advancement have enabled the engineering of tunable synthetic SO coupling in ultracold atomic systems \cite{Spielmann2011_SS, Spielmann2013_SOC, Zhai2015_SOC, SOC_book}.
It is a pertinent issue to investigate the effect of dissipation in such SO coupled systems when its spin is coupled to the environmental degrees of freedom. The non-equilibrium dynamics of particles with SO coupling are directly relevant for understanding transport, magnetic and other physical properties of the materials \cite{spintronics_application, spintronics_review, spintronics_Macdonald, Schaffer2016, Graphene_spintronics1, sdg_dirac, Graphene_spintronics2}.
Furthermore, both intra-particle and environmental entanglement, along with decoherence and dephasing, have emerged as critical issues in quantum information processing and hardware development \cite{Schlosshauer2019, Unruh1995, Chin2014_environmental_entanglement,  Nielson_and_Chuang, Brandt1999, Horodecki_review, Urbasi_review, Azzini_review, Konrad2017, Adhikari2010}.
Additionally, rapid progress in ultracold atomic systems, ion traps, and cavity quantum electrodynamics setups has opened new avenues for realizing the SB model and exploring a range of fascinating physical phenomena \cite{Grifoni2018, Nori2019, Fritz2018, Cirac2008, Sun2025, Duan2024, Plenio2018}. Structured bath can be engineered using trapped ion system, where the spectral density of the bath can be tailored to control dissipation \cite{Sun2025, Duan2024, Plenio2018}. 
Moreover, the dissipative dynamics of a SO coupled impurity interacting with a phonon bath of a condensate can give rise to intriguing forms of quantum Brownian motion \cite{Petruccione_book, Weiss_book, Caldeira_Leggett_1981, QBM_review1, QBM_review2, QBM_SOC1, QBM_SOC2}.

In this work, we consider a particle with one-dimensional (1D) SO interaction which is coupled to a bosonic environment through its spin degree of freedom, describing a SO coupled spin boson model. 
Such tunable SO coupling has been achieved experimentally \cite{Spielmann2011_SS, Spielmann2013_SOC, Zhai2015_SOC, SOC_book}, facilitating studies of rich quantum phases of interacting bosonic condensates, particularly in the realization of the supersolid stripe phase \cite{Spielmann2011_SS, Ketterle2017_SS, Tarruell2024_SS, Zhang2011_SOC, Stringari2015_SOC, Stringari_SS_review, Sinha_SS_review}.
We explore the ground state properties of this dissipative SO coupled system within the framework of variational polaron transformation \cite{Zheng2011}. This system undergoes a magnetization transition as the coupling to the environment increases. 
Notably, the environment induces a significant change in the energy-momentum dispersion relation, wherein the double-degenerate minima at finite momentum collapse into a single minimum at zero momentum as the system-bath coupling exceeds a critical value.
This can have important consequences in transport properties and quantum phases of interacting SO coupled systems \cite{Spielmann2011_SS, Spielmann2013_SOC, Zhai2015_SOC, SOC_book, spintronics_application, spintronics_Macdonald, Schaffer2016, spintronics_review, Graphene_spintronics1, sdg_dirac, Graphene_spintronics2, Ketterle2017_SS, Tarruell2024_SS, Zhang2011_SOC, Stringari2015_SOC}.
Furthermore, the entanglement entropy serves as another indicator of this transition, which also has interesting aspects in the presence of environment, making it relevant for quantum information processing.
In the presence of a weak trapping potential, we also observe such transition, which is accompanied by a sharp jump in magnetization as well as entanglement entropy.
Also, the change in quasi-degenerate ground state is reflected from rapid increase in the ground state energy gap.
This model is relevant for both solid state systems as well as cold atom setups, where the bath can also be engineered.

\section{Model and variational analysis}
We consider a 1D SO coupled particle in the presence of a bosonic bath described by the total
 Hamiltonian, $\hat{\mathcal{H}} = \hat{\mathcal{H}}_{\rm SO} + \hat{\mathcal{H}}_{\rm B} + \hat{\mathcal{H}}_{\rm I}$, where,
\begin{subequations}
\begin{align}
\hat{\mathcal{H}}_{\rm SO} &= \frac{\hat{p}^2_{x}}{2m}  -q\hbar \hat{\sigma}_{x}\frac{\hat{p}_{x}}{m} + \delta \hat{\sigma}_{z}, \label{SOC_ham}\\
\hat{\mathcal{H}}_{\rm B} &= \sum_{l} \hbar \omega_{l}\hat{b}^\dagger_{l}\hat{b}_{l}, \label{sb_ham}\\
\hat{\mathcal{H}}_{\rm I} &=\sum_{l} \uplambda_{l} (\hat{b}_{l} + \hat{b}^\dagger_{l}) \hat{\sigma}_{z}. \label{int_ham}
\end{align}
\end{subequations}
The second term in $\hat{\mathcal{H}}_{\rm SO}$ [Eq.~\eqref{SOC_ham}] corresponds to the SO coupled particle with tunable SO coupling strength $q$, and the third term with $\delta$ represents an additional Rabi coupling \cite{Spielmann2011_SS}.
On the other hand, Eq.~\eqref{sb_ham} represents the bosonic bath, where $\hat{b}_{l}$ are the annihilation operators of the bath modes with energy $\hbar\omega_{l }$. The system-bath interaction is described by Eq.~\eqref{int_ham}, where the $z$-component of the spin is linearly coupled to the bath modes with coupling strength $\uplambda_{l}$ similar to the SB model  \cite{Weiss_book, Petruccione_book, Leggett_review}. In the rest of this paper, we set $\hbar,\,k_{B},\,m =1$, and for simplicity, we set $\delta =0$.

Notably, the momentum is a conserved quantity here, since $[\hat{\mathcal{H}},\hat{p}_{x}]=0$. Consequently, the momentum eigenstate 
$\ket{k}$ of the SO coupled particle is the suitable basis to analyze this model. For a fixed momentum $k$, the above Hamiltonian $\hat{\mathcal{H}}$ reduces to that of the SB model with an effective spin tunneling strength $ qk/m$. Next, we analyze this system for a fixed value of $k$ using the variational framework, a method that has been extensively employed to investigate the properties of the SB model \cite{Zheng2011}.
This approach can also provide the energy-momentum dispersion relation of the SO coupled particle, thereby capturing the effects of the bosonic bath.

\section{Variational Analysis for conserved momentum}
To investigate the ground state properties of SO coupled SB model, we employ a variational polaron transformation method, originally developed for the SB model \cite{Silbey_Harris} and later generalized in subsequent studies \cite{Chin2006, plenio_prl, Zheng2011, Tong2011, Chin2014_environmental_entanglement, Chen2018}.
Following unitary transformation is adopted to derive the effective Hamiltonian for a fixed momentum $k$,
\begin{eqnarray}
\hat{\mathcal{U}} = \hat{\mathcal{U}}_{+}\ket{\uparrow}\!\bra{\uparrow}+ \hat{\mathcal{U}}_{-}\ket{\downarrow}\!\bra{\downarrow}, \label{unitary_transformation}
\end{eqnarray} 
where $\hat{\mathcal{U}}_{\pm} = \exp\left(-\sum_{l}f^{(\pm)}_{l}(\hat{b}_{l}-\hat{b}^\dagger_{l})\right)$ generates displacement by an amount $f^{(\pm)}_{l}$  of the bath oscillators, effectively describing them as coherent states. Here the shifts $f^{(\pm)}_{l}$ are considered as variational parameters. The states $\ket{\uparrow}$ and $\ket{\downarrow}$ denote the spin-up and spin-down states, respectively. By applying the transformation in Eq.~\eqref{unitary_transformation} to the Hamiltonian $\hat{\mathcal{H}}$, we obtain the effective Hamiltonian $\hat{\mathcal{H}}'$ as follows, 
\begin{eqnarray}
\hat{\mathcal{H}}' = \hat{\mathcal{U}}^\dagger \hat{\mathcal{H}} \hat{\mathcal{U}} = \hat{\mathcal{H}}'_{0} + \hat{\mathcal{H}}'_{1} + \hat{\mathcal{H}}'_{2}, \label{transformed_ham}
\end{eqnarray}
where the individual terms for a fixed momentum $k$ are given by,
\begin{subequations}
\begin{align}
\hat{\mathcal{H}}'_{0} =& \,\frac{k^2}{2}\!-\!|k|\tilde{q}\hat{\sigma}_{x} \!+\! B_{+}\ket{\uparrow}\!\bra{\uparrow} \!+\! B_{-}\ket{\downarrow}\!\bra{\downarrow} \!+\! \sum_{l}\omega_{l}\hat{b}^\dagger_{l}\hat{b}_{l},\label{ham_0} \\
\hat{\mathcal{H}}'_{1} =& \sum_{l}(\hat{b}_{l} + \hat{b}^\dagger_{l})\,\Big[\!\left(f^{(+)}_{l}\omega_{l}\!+\! \uplambda_{l}\right)\!\ket{\uparrow}\!\bra{\uparrow}\!+\!\left(f^{(-)}_{l}\omega_{l}\!-\!\uplambda_{l}\right)\!\ket{\downarrow}\!\bra{\downarrow}\!\Big] \notag \\ 
&+\!\dot{\imath}|k|\tilde{q}\hat{\sigma}_{y}\sum_{l}(\hat{b}^\dagger_{l}-\hat{b}_{l})\left(f^{(+)}_{l}-f^{(-)}_{l}\right)\label{ham_1}\\
\hat{\mathcal{H}}'_{2} =& -\!|k|q\hat{\sigma}_{x}\left[\cosh\left(\sum_{l}\left(f^{(+)}_{l}-f^{(-)}_{l}\right)(\hat{b}^\dagger_{l}-\hat{b}_{l})\right)-\eta\right]\notag\\
&+\!\dot{\imath}|k|q\hat{\sigma}_{y}\Bigg[\!\sinh\left(\sum_{l}\left(f^{(+)}_{l}-f^{(-)}_{l}\right)(\hat{b}^\dagger_{l}-\hat{b}_{l})\right)\notag\\
&-\!\eta\!\sum_{l}\left(f^{(+)}_{l}-f^{(-)}_{l}\right)(\hat{b}^\dagger_{l}-\hat{b}_{l})\Bigg]. \label{ham_2}
\end{align}\label{transformed_ham_full}
\end{subequations}
Here, $B_{\pm} \!\!=\!\! \sum_{l}\left(f^{(\pm) 2}_{l}\omega_{l} \pm 2f^{(\pm)}\uplambda_{l}\right)$, and the renormalized SO coupling is represented by $\tilde{q} = q\eta$ with,
\begin{eqnarray}
 \eta = \exp\left(-\frac{1}{2}\sum_{l}\left(f^{(+)}_{l}-f^{(-)}_{l}\right)^2\right),  \label{effective_SOC_param}
\end{eqnarray}
which plays a key role in the delocalization to localization transition in the SB model \cite{Silbey_Harris}.
In the above transformed Hamiltonian, we only consider the first two terms and neglect $\hat{\mathcal{H}}'_{2}$, as it describes two particle and higher order excitation process above the bosonic vacuum state \cite{Zheng2011}.
The Hamiltonian in Eq.~\eqref{ham_0} can be diagonalized, yielding two energy branches.
\begin{eqnarray}
\mathcal{E}_{\pm}(k) = \frac{k^2}{2}+\frac{\left(B_{+}+B_{-}\right)}{2} \pm \frac{1}{2}\sqrt{(B_{+}-B_{-})^2+4k^2\tilde{q}^2}, \label{energy_branches}
\end{eqnarray} 
Correspondingly, the ground state wavefunction can be written in the form,
\begin{eqnarray}
\ket{G} = \left(u_{-}\ket{\uparrow} + v_{-}\ket{\downarrow}\right) \otimes \ket{0}_{b} \otimes  \ket{k}, 
\end{eqnarray}
where $u_{-}(v_{-})$ are given by,
\begin{align}
u_{-} &= \sqrt{\frac{1}{2}\left(1-\frac{\epsilon}{W}\right)},\quad v_{-} = \sqrt{\frac{1}{2}\left(1+\frac{\epsilon}{W}\right)}, \label{gs_params}
\end{align} 
with $\epsilon = B_{+}-B_{-}$ and $W = \sqrt{\epsilon^2 + 4k^2\tilde{q}^2}$. Furthermore, assuming that the first order perturbations of the ground state vanishes, i.e. $\hat{\mathcal{H}}_{1}'\ket{G}=0$, the following conditions are obtained,
\begin{subequations}
\begin{align}
\left(\uplambda_{l} + f^{(+)}_{l}\omega_{l}\right)u_{-} + |k|\tilde{q}\left(f^{(+)}_{l}-f^{(-)}_{l}\right)v_{-} &= 0\\
\left(\uplambda_{l} - f^{(+)}_{l}\omega_{l}\right)v_{-} + |k|\tilde{q}\left(f^{(+)}_{l}-f^{(-)}_{l}\right)u_{-} &= 0.
\end{align}
\end{subequations}
Substituting Eq.~\eqref{gs_params} in the above equations yields the form of the variational parameters,
\begin{eqnarray}
f^{(\pm)}_{l} = \frac{\uplambda_{l}}{\omega_{l}}\frac{\left(2|k|\tilde{q}\epsilon/W\mp\sqrt{\left(1-\epsilon^2/W^2\right)}\omega_{l}\right)}{\left(2|k|\tilde{q}+\sqrt{\left(1-\epsilon^2/W^2\right)}\omega_{l}\right)}. \label{self_consistent_eqn}
\end{eqnarray} 
Note that, the same form of the variational parameters $f^{(\pm)}_{l}$ can also be obtained from minimizing $\mathcal{E}_{-}(k)$.  
Moreover, for the ground state $\ket{G}$, $\hat{\mathcal{H}}'_{2}$ does not contribute to the ground state energy within the first order perturbation and can become important only through higher order corrections.

Since the magnetization of the ground state is defined as, $M = \bra{G}\hat{\sigma}_{z}\ket{G}$, it is easy to see that,
\begin{eqnarray}
M = u^2_{-}-v^2_{-} = -\epsilon/W, \label{magnetization_eqn}
\end{eqnarray}
and $B_{\pm}$ can be written as,
\begin{align}
B_{\pm} =& \sum_{l}\frac{\uplambda^2_{l}}{\omega_{l}}\Big(4k^2\tilde{q}^2M^2-\omega^2_{l}(1-M^2)-4|k|\tilde{q}\omega_{l}\sqrt{1-M^2}\notag\\
&\mp 8k^2\tilde{q}^2M\Big)/\left(2|k|\tilde{q} + \omega_{l}\sqrt{1-M^2}\right)^2. \label{zero_point_terms}
\end{align}
To study the effects of the bath on the ground state properties in a systematic manner, we now consider the following spectral function of the bath \cite{Bulla2005},
\begin{eqnarray}
J(\omega)&=&\pi\sum_{l}\uplambda^2_{l}\delta(\omega-\omega_{l})\notag\\
&=& 2\pi\alpha\omega^{1-s}_{c}\omega^s\Theta(\omega_{c}-\omega), 
\end{eqnarray}
where $0<s<1$, $s=1$, $s>1$, denote a sub-ohmic, ohmic, and super-ohmic bath, respectively. Here, $\omega_{c}$ is the cut-off frequency of the bath and $\alpha$ denotes the coupling constant. For the case of sub-ohmic bath, in the limit $\omega_{c}\!\gg\! kq$, the lowest branch of energy corresponding to the ground state [Eq.~\eqref{energy_branches}] in the leading order of $|k|q/\omega_{c}$ can be written as,
\begin{align}
\mathcal{E}_{-}(k) =&\, \frac{k^2}{2} \!-\! 2\alpha\frac{\omega_{c}}{s} \!+\! 2\alpha\omega_{c}(1+M^2)\frac{\pi(1-s)}{\sin{\pi s}}\!\left(\frac{2|k|\tilde{q}}{\omega_{c}\sqrt{1-M^2}}\right)^{\!s} \notag\\ 
&\!-\!\frac{1}{2}\left[\left(-8M\alpha\omega_{c}\frac{\pi(1-s)}{\sin{\pi s}}\right)^{\!2}\!\!\left(\frac{2|k|\tilde{q}}{\omega_{c}\sqrt{1-M^2}}\right)^{\!2s} \!\!\!\!+ 4k^2\tilde{q}^2\right]^{\!\frac{1}{2}}\!\!, \label{modified_dispersion}
\end{align} 
where the effective SO coupling strength $\tilde{q}$ is given by,
\begin{align}
\tilde{q} =&\, q\exp\left(-4\alpha\!\!\int^{\omega_{c}}_{0}\!\!\frac{\omega^{1-s}_{c}\omega^s(1-M^2)}{\left(2|k|\tilde{q}+\omega\sqrt{1-M^2}\right)^2}d\omega\right)\notag \\
=&\, q\exp\left(-4\alpha\frac{\pi s}{\sin{\pi s}}\left(\frac{2|k|\tilde{q}}{\omega_{c}\sqrt{1-M^2}}\right)^{\!s-1}\right)\exp\left(\frac{4\alpha}{(1-s)}\right). \label{renormalized_momentum}
\end{align}
The above Equation and Eqs.~(\ref{magnetization_eqn},\ref{zero_point_terms}) can be solved self-consistently to yield the ground state energy $\mathcal{E}_{-}(k)$ and magnetization $M$ for a fixed value of momentum $k$.

\begin{figure}
	\centering
	\includegraphics[width=1.02\columnwidth]{Fig1.pdf}
	\caption{{\it Dissipation induced transition}: Changes in the dispersion of the lower energy branch with increasing coupling strength $\alpha$ for sub-ohmic dissipation with $s=0.4$. From the top to the bottom curve: $\alpha = 0.001,\, 0.005,\, 0.010,\, 0.015,\, 0.020,\, {\rm and}\, 0.025$, respectively. The localization transition of the ground state occurs at $\alpha_{c} \sim 0.0187$, and colorscale represents the Magnetization $M$. Colorscale map of the (b) magnetization $M$ and (c) Entanglement entropy $S_{en}$, in the $k$-$\alpha$ plane. (d) Variation of Entanglement entropy $S_{en}$ (left axis, blue curves) and magnetization $M$ (right axis, red curves) with coupling $\alpha$ for fixed values of $k$, denoted by the dashed ($k=0.25$) and dotted lines ($k=1.00$) in (b,c). The green solid vertical lines denote the localization transitions for fixed values of $k$. In all figures, the cut-off frequency is chosen as $\omega_{c} = 10$. In these and the rest of the figures, the energies and momenta are in the units of $q^2$ and $q$, respectively.}
\label{Fig1}
\end{figure}

Importantly, above self-consistent procedure yields the energy-momentum dispersion of the lowest branch, which undergoes significant change due to the coupling with bosonic bath. In the absence of a bath, the dispersion of SO coupled particle $\mathcal{E}_{-}(k) = \frac{k^2}{2} - q|k|$ has two minima at $\pm q$ and its spin is polarized along the x-axis depending on the direction of the momentum $k$, with the wavefunction $\frac{1}{\sqrt{2}}(\ket{\uparrow}+sgn(k)\ket{\downarrow})$.
With increasing the coupling to bath the dispersion gradually changes and most importantly, above a critical coupling strength this lowest energy branch exhibits only one minimum at momentum $k =0$. as shown in Fig.~\ref{Fig1}(a).
Furthermore, the magnetization of the particle also undergoes a continuous transition. This SO coupled particle acquires a magnetized phase with $M \neq 0$ due to the coupling with the bath. To understand this transition for a fixed value of $k$, the energy dispersion in Eq.~\eqref{modified_dispersion} can be expanded in a Landau Ginzburg form, with the magnetization $M$ serving as the order parameter \cite{plenio_prl}, where the coefficients can depend on $k$. Consequently, for a fixed $k$, the SO coupled particle acquires a finite magnetization $M\neq 0$ continuously above a critical coupling strength.

For a given dissipation strength $\alpha$, the magnetization along the $z$-axis decreases continuously starting from $k=0$ and vanishes beyond a critical momentum $|k|>k_c$. The colored regions in the energy momentum dispersion curves of Fig.~\ref{Fig1}(a) indicate finite magnetization of the particle. In Fig.~\ref{Fig1}(b), the magnetization $M$ is represented by the color scale in the $k$-$\alpha$ plane, clearly revealing the phase boundary associated with this continuous magnetic transition.

Next, we investigate the entanglement between spin and the environment for different momentum $k$ of the particle. By integrating out the bath degrees of freedom, the reduced density matrix of the SO coupled particle takes the form ${\rm Tr}_{\rm B} \ket{\psi}\!\bra{\psi} = \hat{\rho}_s \ket{k}\!\bra{k} $, where $\hat{\rho}_s $ denotes the reduced density matrix corresponding to the spin degree of freedom, which is given by,
\begin{equation}
\hat{\rho}_s = |u_{-}|^2\ket{\uparrow}\!\bra{\uparrow} + |v_{-}|^2\ket{\downarrow}\!\bra{\downarrow} + u_{-}v_{-}\eta\left[\,\ket{\uparrow}\!\bra{\downarrow} + \ket{\downarrow}\!\bra{\uparrow}\,\right] 
\end{equation}
The degree of entanglement with the environment can be quantified from the entanglement entropy $S_{en} = -{\rm Tr}(\hat{\rho}_s {\rm ln} \rho_s)$ which has implicit dependence on the momentum $k$ of the particle. The entanglement entropy increases from $k=0$, attains a maximum at a critical momentum corresponding to the magnetic phase transition, and then decreases monotonically with $k$ [see Fig.~\ref{Fig1}(d)]. As shown in Fig.~\ref{Fig1}(c), the entropy reaches its peak at the boundary of the magnetized phase in the $k$-$\alpha$ plane, thereby capturing the continuous nature of the transition. 
Note that, the sharp change in the magnetization across the critical coupling can get smeared in the presence of a small non-vanishing bias ($\delta \neq 0$), which has been observed in the usual SB model \cite{Zheng2011}.

\begin{figure}
	\centering
	\includegraphics[width=1.02\columnwidth]{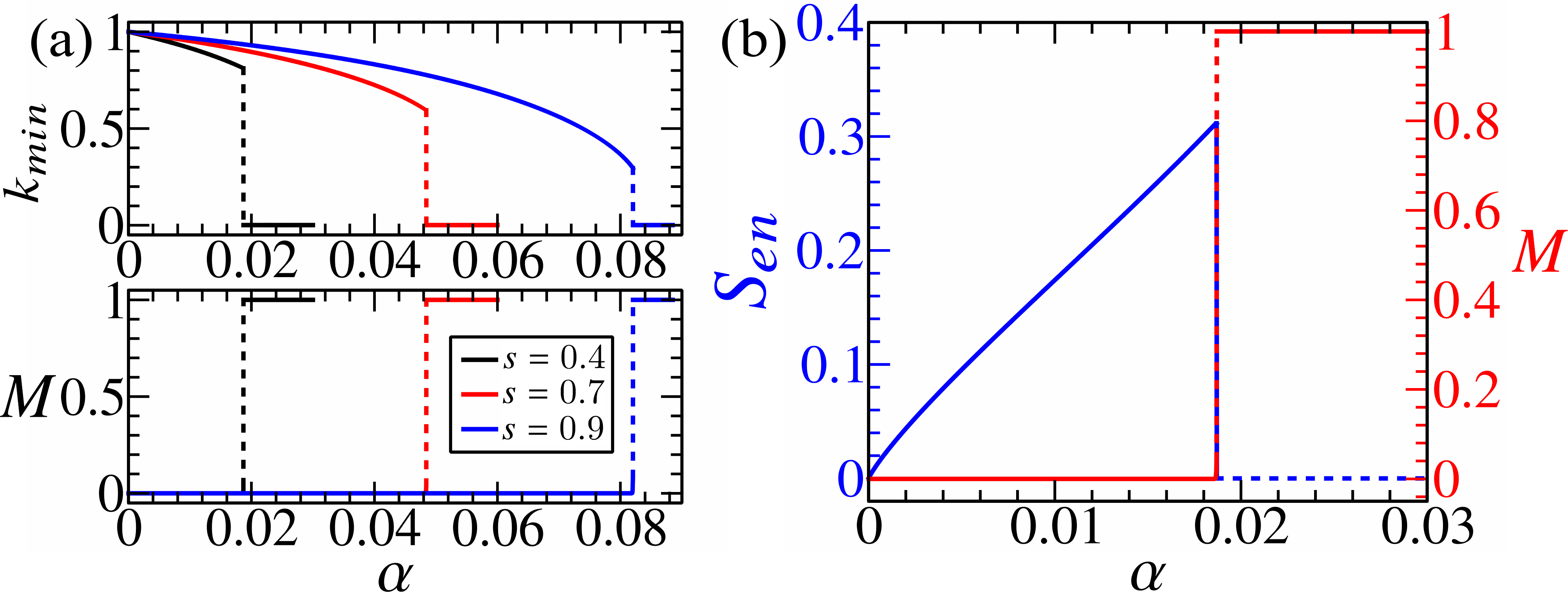}
	\caption{{\it Ground state properties}: (a) Variation of momentum $k_{min}$ (top) and Magnetization $M$ (bottom) corresponding to the minimum energy state with increasing $\alpha$ for different values of $s$. (b) Variation of the entanglement entropy $S_{en}$ (left axis) and Magnetization $M$ (right axis) corresponding to the ground state energy with increasing $\alpha$ for $s=0.4$. The jump at $\alpha_{c}$ indicates a first-order transition. In all figures, the cut-off frequency is chosen as $\omega_{c} = 10$.}     
\label{Fig2}
\end{figure}

We also focus on the minima of the energy dispersion $\mathcal{E}_{-}(k)$, which in absence of bath appear at $k = \pm q$, representing doubly degenerate ground states. In the presence of bosonic bath, the strength of SO interaction $q$ is renormalized, resulting in  a decrease in momentum at which the doubly degenerate minima appear. As depicted in the top panel of Fig.~\ref{Fig2}(a), with increasing the coupling strength $\alpha$, the momentum $k_{min}$ corresponding to the doubly degenerate ground states decreases and sharply vanishes above a critical coupling strength $\alpha_c$, indicating single minimum of the dispersion relation $\mathcal{E}_{-}(k)$ at $k=0$. This indicates a sharp first order transition of the ground state of the SO coupled particle as the coupling with the bosonic bath is increased. Moreover, both magnetization and entanglement entropy exhibit discontinuous jumps across this transition [see Fig.~\ref{Fig2}].
This transition also depends on the spectral density of the bath. As evident in Fig.~\ref{Fig2}(a), the critical coupling corresponding to this transition increases with increasing values of $s$ characterizing the sub-ohmic bath. The discontinuous jump in the momentum and magnetization also decreases with increasing $s$. 
Such a change in the spectrum can also be understood from the effective single-particle dispersion relation given in Eq.~\eqref{energy_branches}. Moreover, the curvature at the minima physically correspond to the effective mass of the particle
$\frac{1}{m^*}= \frac{\partial^2 \mathcal{E}}{\partial k^2}$.
Below the transition, the effective mass corresponding to the minima at $k_{min}$ can be estimated by $m^* \simeq 1/(1 - \epsilon^2 /4\tilde{q}^4)$. A single minimum at $k=0$ emerges when $|\epsilon| > 2\tilde{q}^2$, where the effective mass can take the form $m^* \simeq 1/(1 - 2\tilde{q}^2/|\epsilon|)$. This analysis indicates that the generation of an effective magnetic field $\epsilon$, arising from the asymmetric displacement of bath oscillators, plays a crucial role in determining the shift of the energy minimum. Moreover, the modification of the effective mass induced by the environment represents an intriguing manifestation of dissipation, akin to the polaronic mass renormalization experienced by a mobile impurity interacting with phonons \cite{Demler2016,Demler2017,Demler_review}.

Here we emphasize that, two distinct types of transitions can occur in the SO coupled SB model: (i) for a fixed value of $k$ (or $\alpha$), the magnetization $M$ exhibits a continuous transition with increasing $\alpha$ (or $k$); and (ii) a discontinuous transition of the ground state with increasing coupling $\alpha$. Although, the effective dispersion of the lowest branch given in Eq.~\eqref{modified_dispersion} plays a central role in understanding both the transitions,  determining the ground state, as mentioned previously, requires a complicated minimization of the energy subjected to self-consistent conditions.
	
The discontinuous transition of the ground state can clearly be understood for SO coupled particle in the presence of a harmonic trap which we discuss in the next section.

\section{Variational analysis in presence of Harmonic trap}
In this section, we investigate how the ground state of the SO coupled system in the presence of a harmonic trap modifies due to interaction with the bosonic bath modes.
In the presence of a harmonic trap (with frequency $\omega_{0}$), the SO coupled Hamiltonian in Eq.~\eqref{SOC_ham} can be written as,
\begin{eqnarray}
\hat{\mathcal{H}}_{\rm SO}  &=&   \dot{\iota} q\sqrt{\omega_{0}/2}\, (\hat{a}-\hat{a}^\dagger)\hat{\sigma}_{x} + \omega_{0}\hat{a}^\dagger \hat{a}, \label{SOC_ham_trap}
\end{eqnarray}
where $\hat{a}=(\hat{x}\sqrt{\omega_{0}/2}+\dot{\imath}\hat{p}_{x}/\sqrt{2\omega_{0}})$ denotes the ladder operator. Note that, unlike the previous case, momentum is not a conserved quantity any more in the presence of the trap. It is evident from Eq.~\eqref{SOC_ham_trap}, there are doubly degenerate ground states, given by,$\ket{\mathcal{U}_{\pm q}} \otimes \frac{1}{\sqrt{2}}[\ket{\uparrow} \pm \ket{\downarrow}]$, where
\begin{equation}
\ket{\mathcal{U}_{\pm q}} = \exp\left(\frac{\pm\dot{\imath}q}{\sqrt{2\omega_{0}}}(\hat{a}+\hat{a}^\dagger)\right)\ket{0}_{a}.
\label{mom_coherent}
\end{equation}
with $\ket{0}_a $ representing the vacuum state of the oscillator.
These states correspond to the coherent states with momentum $\pm q$, and the spin being aligned along the $\pm x$-axis, respectively.  Due to the presence of the bosonic bath modes, an effective magnetic field $\epsilon=B_{+}-B_{-}$ is generated in the single-particle Hamiltonian [see Eq.~\eqref{ham_0}], which tries to polarize the spins along the $z$-axis. In the presence of an additional polarising term $\epsilon \sigma_z$, the degenerate ground states can be written as,
\begin{equation}
\ket{\phi^{0}_{\pm k}} = \left[\cos{\left(\frac{\theta}{2}\right)}\ket{\uparrow} \pm \sin{\left(\frac{\theta}{2}\right)}\ket{\downarrow}\right]\ket{\mathcal{U}_{\pm k}},
\label{deg_nigs}
\end{equation}
where,the spin orientations, expressed in terms of the $SU(2)$ representation, describe a symmetric arrangement of spin vectors along the positive and negative $x$-axes, each making an angle $\theta$ with the $z$-axis.

Based on the above observations, we can construct a variational wavefunction in the presence of bath using the polaron transformation and a linear combination of these degenerate states with equal and opposite momentum $k$, that is given by, 
\begin{equation}
|\psi_G\rangle = \frac{1}{\sqrt{\mathcal{N}}}[\cos{\!\chi}\ket{\phi_{+k}} + \sin{\!\chi}\ket{\phi_{-k}}],
\label{var_wf_trap}
\end{equation}
where,
\begin{equation}
\ket{\phi_{\pm k}} = \left[\cos{\left(\frac{\theta}{2}\right)}\ket{\uparrow}\ket{\mathcal{B}_{+}} \pm \sin{\left(\frac{\theta}{2}\right)}\ket{\downarrow}\ket{\mathcal{B}_{-}}\right]\ket{\mathcal{U}_{\pm k}}.
\label{coherent}
\end{equation}
%
%
Here, $\chi$ denotes the mixing between the two coherent states with equal and opposite momentum $k$, and $\mathcal{N}\!\!=\!\!1 + \sin{(2\chi)}\cos{\theta}e^{-k^2l^2}$ is the normalization constant, where $l\!\!=\!\!\sqrt{1/\omega_{0}}$ is the trap confinement length.
As discussed previously, the shifted bath oscillators are denoted by $\ket{\mathcal{B_{\pm}}} \!\!=\!\! \hat{\mathcal{U}}_{\pm}\ket{0}_{b}$. Here we consider $\chi$, $\theta$, $f^{(\pm)}_{l}$, and $k$ as the variational parameters. We minimize the ground state energy $E \!=\! \bra{\psi_G}\hat{\mathcal{H}}\ket{\psi_G} $ with respect to the variational parameters in order to explore the ground state properties, particularly how the magnetization and momentum $k$ change as the coupling with bath increases.

\begin{figure}
	\centering
	\includegraphics[width=1.02\columnwidth]{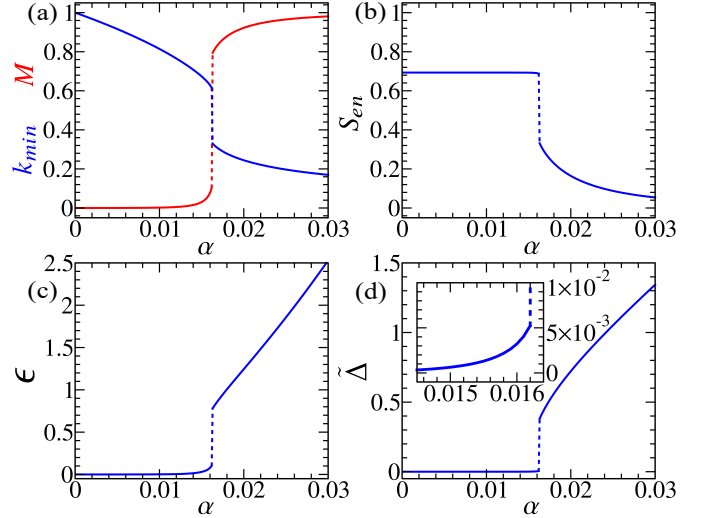}
	\caption{{\it Ground state properties in the presence of a harmonic trap}: (a) Variation of the momentum $k_{min}$ (blue) and Magnetization $M$ (red) corresponding to the minimum energy state with increasing $\alpha$. (b) Variation of the entanglement entropy $S_{en}$  corresponding to the ground state with increasing $\alpha$. Variation of (c) effective magnetization $\epsilon = B_{+}-B_{-}$ and (d) splitting energy $\tilde{\Delta} = 2|\Delta|e^{-k^2l^2}/\mathcal{N}$, with increasing $\alpha$. The inset in (d) indicates a zoom region near $\alpha=\alpha_{c}$. The jump at $\alpha_{c}\sim 0.0162$ indicates a first-order transition. Parameters chosen: $s=0.4$, $\omega_{0}=0.1$, and $\omega_{c} = 10$.}     
\label{Fig3}
\end{figure}

For the sub-ohmic bath ($0\!\!<\!\!s\!\!<\!\!1$), in the limit $\omega_{c} \!\gg\! kq$, the energy corresponding to the variational wavefunction [Eq.~\eqref{var_wf_trap}] in the leading order of $kq/\omega_{c}$ is given by,
\begin{align}
\mathcal{E}_{-}(k) =&\, \frac{k^2}{2\mathcal{N}}\left(1-\sin{(2\chi)}\cos{\theta}e^{-k^2l^2}\right) - \frac{k\tilde{q}}{\mathcal{N}}\sin{\theta} - 2\alpha\frac{\omega_{c}}{\mathcal{N}s}(\mathcal{C}_{+} \notag\\
+&\mathcal{C}_{-}) + 2\alpha\frac{\omega_{c}}{\mathcal{N}}\frac{\pi(1-s)}{\sin{(\pi s)}}\frac{4 \,\mathcal{C}_{+}\mathcal{C}_{-}}{(\mathcal{C}_{+}+\mathcal{C}_{-})}\left(\frac{k\tilde{q}\,(\mathcal{C}_{+}+\mathcal{C}_{-})\sin{\theta}}{2\,\mathcal{C}_{+}\mathcal{C}_{-}\omega_{c}}\right)^s,
\end{align}
where $\mathcal{C}_{\pm}\!=\widetilde{\mathcal{C}}_{\pm}\left(1\pm\sin{(2\chi)}e^{-k^2l^2}\right)$, with $\widetilde{\mathcal{C}}_{+} \!\!= \!\cos^2\!{(\theta/2)}$ and $\widetilde{\mathcal{C}}_{-}\!\! =\! \sin^2\!{(\theta/2)}$, respectively. Minimizing the above energy expression with respect to the variational parameters in a self-consistent manner provides an estimate of both energy and the structure of the ground state. We obtain the magnetization $M \!=\! \bra{\psi} \sigma_z \ket{\psi}$ and momentum of the coherent states $k\!=\!k_{min}$ corresponding to the minimum energy state. Their variation with the bath coupling strength $\alpha$ is depicted in Fig.~\ref{Fig3}(a), both of which exhibit a sharp jump at a critical coupling, signalling a first-order transition. As expected from the degenerate ground states of non-interacting SO coupled particle, for weak coupling with the bath, we find $k_{min}\!\approx\!q$, while the magnetization remains vanishingly small. On the other hand, the momentum $k_{min}$ drops significantly and magnetization $M$ approaches to unity above the transition. In the absence of the system-bath coupling, any linear combination of the degenerate states $\ket{\phi_{+k}}$ and $\ket{\phi_{-k}}$ is possible, however, for non-vanishing coupling, we obtain a symmetric ground state with equal mixing, for which $\chi=\pi/4$ minimizes the energy.

Next, we analyze the entanglement properties of the spin sector from the corresponding reduced density matrix $\rho_s$, given by,
\begin{eqnarray}
\rho_s &=& \mathrm{Tr_{B}}\!\! \int\!\! dx \langle x |\psi_G \rangle \langle \psi_G|x \rangle,\notag\\
&=& \frac{\mathcal{C}_{+}}{\mathcal{N}}\ket{\uparrow}\!\bra{\uparrow} + \frac{\mathcal{C}_{-}}{\mathcal{N}}\ket{\downarrow}\!\bra{\downarrow}\notag\\ 
&&+ \frac{\eta}{2\mathcal{N}}\cos{(2\chi)}\sin{\theta}\left[\,\ket{\uparrow}\!\bra{\downarrow}+\ket{\downarrow}\!\bra{\uparrow}\,\right]
\label{rho_trap}
\end{eqnarray}
where, $\mathrm{Tr_{B}}$ represents tracing over the bath oscillators. We evaluate the entanglement entropy of the spin sector $S_{en} \!\!=\! -\mathrm{Tr}( \rho_s {\rm ln}\rho_s)$ and study its variation with bath coupling strength $\alpha$. Stark contrast in entanglement behavior is observed due to the presence of a trap, compared to that of free SO coupled particle discussed in the previous section. Unlike the free case momentum is no longer a conserved quantity in the presence of a potential and the ground state $\ket{\psi_G}$ becomes a symmetric combination of two equal and opposite momentum carrying states with spins aligned in opposite direction along the $x$-axis [see Eq.~\eqref{coherent}]. 
Consequently, below the transition ($\alpha \!<\! \alpha_c$), the spin sector exhibits strong entanglement, as reflected in the reduced density matrix $\rho_s$, which approaches to a maximally mixed state for $k_{min} l \!\gg\! 1$, yielding the maximal entanglement entropy $S_{en}\!\approx\!\ln 2$. While for $\alpha \!>\! \alpha_c $, the entanglement entropy decreases monotonically as $k_{min}l \!\ll\! 1$

In absence of the bath, $\alpha \!\approx\! 0$, the ground state $\ket{\psi_G}$ becomes an arbitrary linear combination of the degenerate states $\ket{\phi^{0}_{\pm k}}$. However for finite coupling with the bath oscillators, the ground state becomes a symmetric combination of opposite momentum carrying states $|\psi_G\rangle \!=\! \frac{1}{\sqrt{\mathcal{N}}}[\ket{\phi_{k}} + \ket{\phi_{-k}}]$. Consequently, the antisymmetric combination of these states gives the excited state $  |\psi_{ex}\rangle \!=\! \frac{1}{\sqrt{\mathcal{N}}}[\ket{\phi_{k}}- \ket{\phi_{-k}}]$. 
To analyze the quasi-degeneracy associated with the ground state we calculate the energy gap,
\begin{equation}
\Delta = \bra{\psi_{ex}}\hat{\mathcal{H}} \ket{\psi_{ex}} - \bra{\psi_{G}}\hat{\mathcal{H}} \ket{\psi_{G}}.
\label{Delta}
\end{equation}
and investigate its variation with the bath coupling $\alpha$.
For $\omega_c \!\!\gg\!\! q^2$, the energy splitting between these two quasi-degenerate states is obtained analytically, and takes the form $\tilde{\Delta} \sim 2|\Delta|e^{-k^2 l^2}\!/\mathcal{N}$, where $\Delta$ can be expressed as,
\begin{eqnarray}
\Delta = -\frac{k^2}{2}\cos{\theta}+B_{+}\cos^2{\!\left(\frac{\theta}{2}\right)}+B_{-}\sin^2{\!\left(\frac{\theta}{2}\right)},
\end{eqnarray}
and $B_{\pm} \!=\! -2\alpha\frac{\omega_{c}}{s} + 2\alpha\omega_{c}\frac{\pi (1-s)}{\sin{(\pi s)}}\frac{4\mathcal{C}^2_{\mp}}{(\mathcal{C}_{+}+\mathcal{C}_{-})^2}\left(\frac{k\tilde{q}\,(\mathcal{C}_{+}+\mathcal{C}_{-})\sin{\theta}}{2\,\mathcal{C}_{+}\mathcal{C}_{-}\omega_{c}}\right)^s$.
Below the critical coupling with the bath, $\alpha <\alpha_c$, and for $ql\gg 1$ corresponding to weak trapping potential, the energy gap is exponentially suppressed as $\sim e^{-k_{min}^2 l^2}$, which reveals the quasi-degeneracy between the states $\ket{\psi_G}$ and $\ket{\psi_{ex}}$. However, the gap $\tilde{\Delta}$ rapidly increases across the transition, signifying the lifting of this quasi-degeneracy as $k_{min}l \ll 1$ [see Fig~\ref{Fig3}(d)]. This phenomenon can also be understood from the generation of an effective magnetic field $\epsilon$ due to the bath, which polarize the spin along the $z$-axis and lifts the degeneracy of the ground state. As shown in Fig~\ref{Fig3}(c), a rapid increase in the polarizing field also occurs across the transition, clearly indicating the link between spin polarization and lifting of ground state degeneracy associated with this transition.
This behavior parallels the transformation of the spectrum $\mathcal{E}(k)$ of the SO coupled particle from a double-minima to a single-minimum structure in the absence of a trap, as discussed previously.

\begin{figure}
	\centering
	\includegraphics[width=1.02\columnwidth]{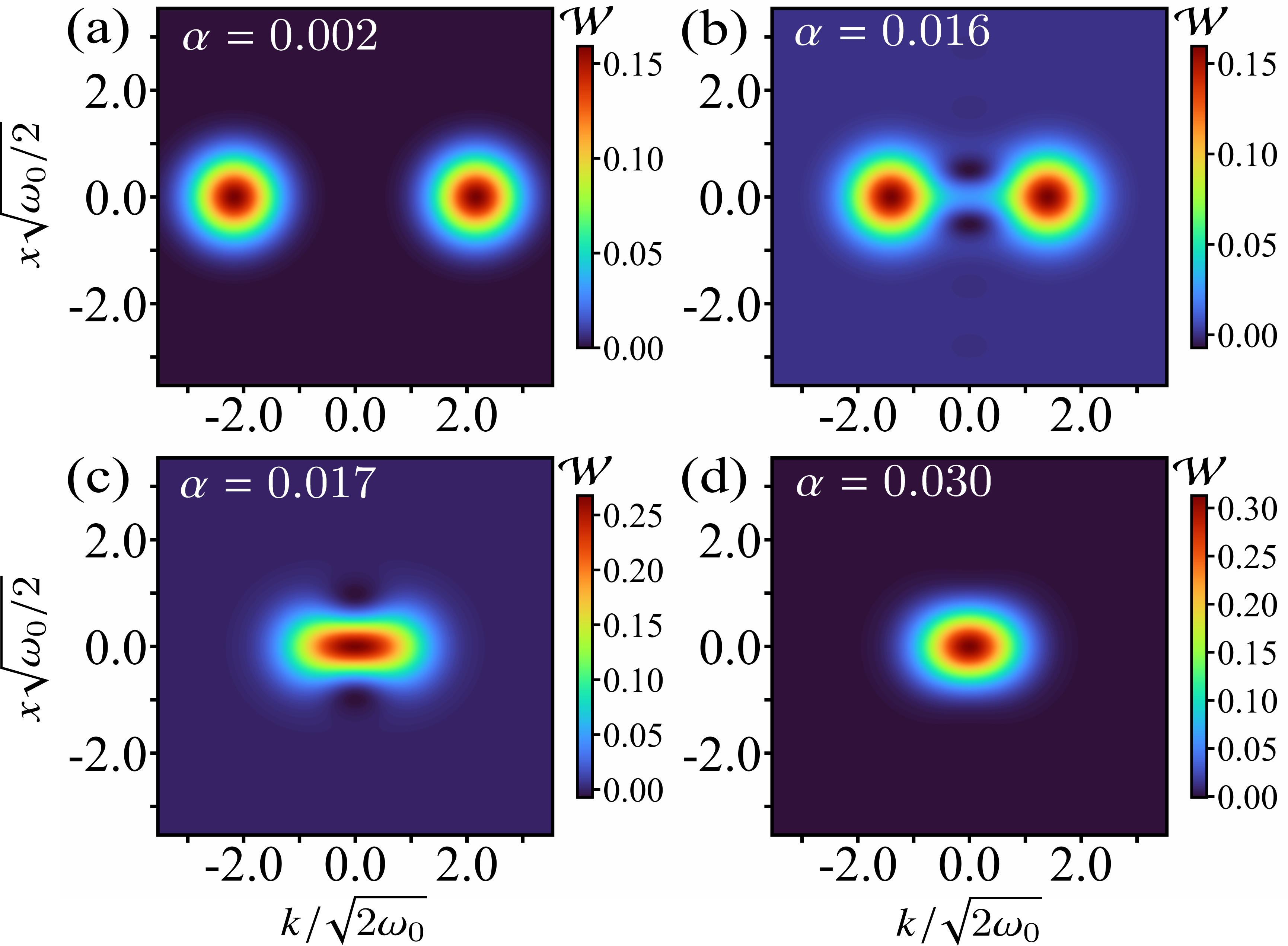}
	\caption{{\it Wigner distributions of the ground state}: Colorscale plot of $\mathcal{W}(\tilde{x},\tilde{k})$ at different values of coupling constant $\alpha$, (a,b) before, and (c,d) after the point of transition, $\alpha_{c} \sim 0.0162$. Parameters chosen: $s=0.4$, $\omega_{0}=0.1$, and $\omega_{c} = 10$.}     
\label{Fig4}
\end{figure}

This change in the ground-state structure with increasing bath coupling can be directly visualized through the semi-classical phase-space density, such as the Wigner distribution.
By tracing out the spin and the bosonic degrees of freedom, we obtain the reduced density matrix $\hat{\rho}_{r} = {\rm Tr}_{B \,S}{\ket{\psi}\!\bra{\psi}}$ of the ground state, which can be written as follows,
\begin{eqnarray}
\hat{\rho}_{r} &=& \frac{1}{\mathcal{N}}\bigg[\frac{1}{2}\big(\ket{\mathcal{U}_{+k}}\!\bra{\mathcal{U}_{+k}}+\ket{\mathcal{U}_{-k}}\!\bra{\mathcal{U}_{-k}}\big)\notag\\
&&+\cos{\theta}\big(\ket{\mathcal{U}_{-k}}\!\bra{\mathcal{U}_{+k}}+\ket{\mathcal{U}_{+k}}\!\bra{\mathcal{U}_{-k}}\big)\bigg]
\end{eqnarray}
The Wigner distribution corresponding to the reduced density matrix is given by,
\begin{eqnarray}
\mathcal{W}(\tilde{x},\tilde{k}) = \frac{1}{\pi}\!\int^{\infty}_{-\infty}\!\!\!\!\!\bra{\tilde{x}-y}\hat{\rho}_{r}\!\ket{\tilde{x}+y}e^{-2\dot{\iota}\tilde{k}y}dy
\end{eqnarray}
where $\tilde{x}=x\sqrt{\omega_{0}/2}$ and $\tilde{k}=k/\sqrt{2\omega_{0}}$ denote the dimensionless position and momentum variables.  
As evident from Fig.~\ref{Fig4}(a), in the weak-coupling regime, the equal superposition of two degenerate momentum-carrying states produces a phase-space density localized around two opposite momenta, resembling a `cat state' like structure, while the interference fringes are suppressed due to the reduced overlap between the spin components. 
However, close to the critical coupling strength, the increase in the effective magnetic field $\epsilon$ enhances the overlap between the two momentum-carrying states, resulting in a non-monotonic phase-space density pattern around $k = 0$ [see Fig.~\ref{Fig4}(b)]. Finally, above the critical coupling, the phase-space density becomes concentrated near zero momentum  [see Fig.~\ref{Fig4}(c,d)], indicating a change in the ground-state structure similar to that of the free-particle case.

\section{Discussions}
We present a spin-orbit (SO) coupled spin-boson (SB) model to explore the effect of dissipation on a particle with SO interaction, for which the spin degree of freedom is coupled to a sub-ohmic heat bath. This system exhibits two distinct types of transitions: a continuous magnetization transition for a free particle with conserved linear momentum, and a discontinuous first-order transition of the ground state as the coupling to the bath increases. Moreover, this transition is accompanied by a striking modification in the energy–momentum dispersion of the SO coupled particle, where the doubly degenerate minima in the spectrum merge into a single minimum at zero momentum with increasing coupling to the bath. From a physical standpoint, the asymmetric displacement of bath oscillators induce an effective magnetic field that plays a crucial role in driving the magnetization transition and lifting the ground-state degeneracy. Moreover, the bath induced effective mass of the particle changes differently across the transition. Such dissipation-driven modifications in the spectrum and magnetic transitions can have a significant impact on transport properties, particularly in spintronic systems \cite{spintronics_application, spintronics_review, spintronics_Macdonald, Schaffer2016, Graphene_spintronics1, sdg_dirac, Graphene_spintronics2}.

Furthermore, we investigate the entanglement properties of the spin sector, where the entanglement entropy exhibits a sharp peak at the boundary of the magnetized phase, capturing the continuous transition, in contrast to the abrupt jump observed across the first-order transition associated with the change in the ground state of free system.
The presence of a weak trapping potential leads to a superposition of the two degenerate ground states with equal and opposite momenta, rendering them quasi-degenerate with an exponentially small energy gap. Importantly, this generates maximally mixed spin state (with high entanglement) below the transition. However, above a critical bath coupling strength, the quasi-degeneracy associated with the ground state is lifted and entanglement decreases monotonically. 
Insights into the behavior of such intra-particle entanglement under environmental influence are highly relevant for quantum information processing \cite{Horodecki_review, Urbasi_review, Azzini_review, Konrad2017, Adhikari2010}. 
Moreover, the superposition of two degenerate states with equal and opposite momenta generates a `cat-like' non-classical state, with phase-space density concentrated around the two opposite momenta, while the interference effects are suppressed due to the orthogonal alignment of the spins, resulting in a dephased cat state.
The study of single-particle physics provides important understanding of exotic quantum phases that emerge from the interplay of SO coupling and inter-particle interactions, which has recently been explored in SO coupled condensates.
In particular, the superposition of two momentum-carrying states gives rise to an intriguing superstripe-supersolid phase, which has been observed experimentally \cite{Spielmann2011_SS, Ketterle2017_SS, Tarruell2024_SS} and has attracted considerable interest in recent years \cite{Zhang2011_SOC, Stringari2015_SOC, Stringari_SS_review, Sinha_SS_review}. It is a pertinent issue to investigate the effect of dissipation on such exotic phases, as it can lead to dephasing of the superstripe phase. 

In the present work, we have used a simple variational analysis to study the basic features of the ground state and its transitions in  a SO coupled system in the presence of dissipation.
Although such a variational analysis can capture the nature of the transition in the spin-boson model specifically for the sub-ohmic bath \cite{Zheng2011}, a more refined treatment may be required to fully account for the approximations used in the present model.
It would also be interesting to investigate the nature of this transition in the presence of baths with different spectral densities. Specifically, understanding the critical behavior may require more sophisticated analytical or numerical techniques as the ohmic limit is approached \cite{Vojta2005, Hou2010}. Recently, SB models with structured spectral densities have been realized in trapped-ion setups \cite{Sun2025, Duan2024, Plenio2018}, providing a promising platform to explore SO coupled SB models with engineered environments.
Although, the bias term is not considered in the present analysis, it would be interesting to study the change in the dispersion as well as the ground state transition of such dissipative SO coupled particle by tuning the bias.

In conclusion, we have analyzed a SO coupled spin-boson system that exhibits a dissipation-driven transition accompanied by a modification of the energy spectrum.
Such effects are relevant to electronic systems with strong SO coupling, including Graphene, as well as to cold-atom setups with tunable synthetic SO coupling, which can also find potential applications in quantum information processing.

\section*{Acknowledgment}
SD is grateful to the Indian National Science Academy for support through its Honorary Scientist Scheme.

\end{document}